\documentclass[12pt,preprint]{aastex}
\begin{document}

\def\lya{Lyman-$\alpha$}  
\def\eqw{\hbox{EW}}
\def\erg{\hbox{erg}}
\def\cm{\hbox{cm}}
\def\sec{\hbox{s}}
\def\ha0{$6559$\AA}
\def\h16{$6730$\AA}
\def\f17{f_{17}}
\def\Mpc{\hbox{Mpc}}
\def\km{\hbox{km}}
\def\kms{\hbox{km s$^{-1}$}}
\def\year{\hbox{yr}}
\def\deg{\hbox{deg}}
\def\msun{M_{\odot}}

\def\ergcm2s{\ifmmode {\rm\,erg\,cm^{-2}\,s^{-1}}\else
                ${\rm\,ergs\,cm^{-2}\,s^{-1}}$\fi}
\def\kmsMpc{\ifmmode {\rm\,km\,s^{-1}\,Mpc^{-1}}\else
   ${\rm\,km\,s^{-1}\,Mpc^{-1}}$\fi}
\def\nv{\ion{N}{5} $\lambda$1240}
\def\civ{\ion{C}{4} $\lambda$1549}
\def\oii{[\ion{O}{2}] $\lambda$3727}
\def\oiipair{[\ion{O}{2}] $\lambda \lambda$3726,3729}
\def\oiii{[\ion{O}{3}] $\lambda$5007}
\def\oiiipair{[\ion{O}{3}] $\lambda \lambda$4959,5007}

\def\kms{{\rm km/s}}
\def\cc{{\rm cm ^{-3}}}
\def\mum{\mu {\rm m}}
\def\kpc{{\rm kpc}}
\def\deg{^\circ}
\def\hpc{{\rm h}^{-1}\,{\rm pc}}
\def\lsun{{\,L_\odot}}

\title{Large equivalent width Lyman-$\alpha$ line emission at z=4.5: young galaxies in a young universe?}

\author{Sangeeta Malhotra \altaffilmark{1,2,3} \and
James E. Rhoads \altaffilmark{3} }

\begin{abstract}
The Large Area Lyman Alpha survey has found $\approx$ 150 \lya\ emitters at
z=4.5. While stellar models predict a maximum \lya\ equivalent width
(EW) of 240 \AA\, 60\% of the \lya\ emitters have EWs exceeding this
value. We attempt to model the observed EW distribution by combining
stellar population models with an extrapolation of Lyman break galaxy
luminosity function at z=4, incorporating observational selection
effects and Malmquist bias. To reproduce the high EWs seen in the
sample we need to postulate a stellar initial mass function (IMF) with
extreme slope $\alpha = 0.5$ (instead of 2.35); zero metallicity
stars; or narrow-lined active galactic nuclei. The models also reveal
that only 7.5-15\% of galaxies need show \lya\ emission to explain the
observed number counts.  This raises the possibility that either
star-formation in high redshift galaxies is episodic or the \lya\
galaxies we are seeing are the youngest 7.5-15\% and that \lya\ is
strongly quenched by dust at about $10^7$ years of age.
\end{abstract}

\altaffiltext{1}{Johns Hopkins University, Charles and 34th Street, Bloomberg center, Baltimore, MD 21218}
\altaffiltext{2}{Hubble Fellow} 
\altaffiltext{3}{Space Telescope Science Institute, 3700 San Martin Drive, Baltimore, MD 21218} 


\section {Introduction}

Nascent galaxies undergoing their first major burst of star formation
are expected to contain many hot, young, massive stars, which ionize
interstellar gas.  Under standard conditions (case B), 
about two \lya\ photons are produced for every three
ionizing photons from the stars. Thus \lya\ photons can be a prominent
signpost of primordial galaxies in formation (Partridge \& Peebles
1967, hereafter PP67). 

Based on these predictions a number of surveys have been carried out
for about three decades (see Pritchet 1994 for a review), with little
success until recently. We have recently undertaken a Large Area Lyman
Alpha survey (Rhoads et al 2000) to identify a large sample of \lya\
emitting galaxies at high redshift ($z=4.5\pm 0.1$) through narrowband
imaging, and so characterize empirically star and galaxy formation in
the early universe.  Both our survey (Rhoads et al. 2000) and other
recent searches over smaller volumes (Cowie \& Hu 1998, Hu et al 1998,
Kudritzki et al 2000, Fynbo \& Moller 2001...others...Pentericci et
al. 2001, Stiavelli et al. 2001) are indeed finding high redshift
\lya\ sources, albeit factors of $\sim 100$ fainter than the early
predictions (Partridge \& Peebles 1967).  This faintness is consistent
with the 30 years of nondetections (1967-1997) by many groups that
preceded the current generation of searches. 

Low luminosity of this line could be due to smaller masses of the
sources, lower star-formation rates, or selective obscuration of
the resonantly scattered \lya\ line (Charlot \& Fall 1993). The latter 
scenario would lead to smaller equivalent width (EW) of the line.
The EW of the \lya\ line can also, in principle, be used as a
diagnostic to determine the population of young, massive stars.
However, this is complicated by several factors: Ionizing flux
depends on stellar metallicity; active galactic nuclei may contribute
to the ionizing flux; and resonant scattering of the \lya\ line
can make it more sensitive to dust than the adjoining continuum, but
(depending on clumping and winds in the interstellar medium) need not do so
(Neufeld 1991; Kunth et al 1998). 

In this paper we report on the distribution of equivalent widths of
\lya\ line for sources detected in the LALA survey. In section 2.1, we
will discuss briefly the LALA survey and selection critera for
sources.  Section 2.2 describes how we calculate the equivalent
width of the \lya\ line and presents the observed distribution.
Section 3 contains details of stellar population modelling,
and section 4 contains discussions and conclusions.

\section {Observations and equivalent widths}

\subsection{Survey and data reduction}
An efficient search for \lya\ emitters (and other emission line
galaxies) was started in 1998 using the CCD Mosaic Camera at the Kitt
Peak National Observatory's 4m Mayall telescope. The Mosaic camera has
eight $2048 \times 4096$ chips in a $4 \times 2$ array comprising a
$36'\times 36'$ field of view. The final area covered by the LALA
survey is $0.72$ square-degrees in two MOSAIC fields centered at
14:25:57 +35:32 (2000.0) and 02:05:20 -04:55 (2000.0).
Five overlapping narrow band filters of width FWHM$\approx 80$\AA\ 
are used. The central wavelengths are $6559$, $6611$, $6650$,
$6692$, and $6730 $\AA, giving a total redshift coverage $4.37 < z <
4.57$.  This translates into surveyed comoving volume of $7.4 \times
10^5$ comoving $\Mpc^3$ per field for $H_0 = 70
\kmsMpc$, $\Omega_M = 0.3$, $\Omega_\Lambda=0.7$. 

In this paper we report on the data from the spring field centered at
14:25:57 +35:32 (2000.0), since all the imaging on this field at $z
\approx 4.5$ is complete. In about 6 hours per filter per field we
achieve line detections of about $2 \times 10^{-17} \ergcm2s$. The
survey sensitivity varies with seeing.  Broadband images of these
fields in a custom $B_w$ filter ($\lambda_0 = 4135$\AA,
$\hbox{FWHM}=1278 $\AA) and the Johnson-Cousins $R$, $I$, and $K$
bands were taken as part of the NOAO Deep Widefield Survey
(Jannuzi \& Dey 1999).

The images were reduced using the MSCRED package (Valdes \& Tody 1998;
Valdes 1998) in the IRAF environment (Tody 1986, 1993), together with
assorted custom IRAF scripts. Details of the data reduction can be
found in the paper by Rhoads et al (2000).  Catalogs were generated
using the SExtractor package. 
Fluxes were measured in $2.32''$ (9 pixel) diameter apertures, and
colors were obtained using matched $2.32''$ apertures in registered
images that had been convolved to yield matched point spread
functions. In order to gracefully handle sources that are not detected
in all filters, we have chosen to use ``asinh magnitudes'' (Lupton,
Gunn, \& Szalay 1999), which are a logarithmic function of flux for
well detected sources but approach a linear function of flux for weak
or undetected sources.  The color scatter achieved for bright sources
($R<22$) is $0.10$ magnitudes (semi-interquartile range).  This
includes the true scatter in object colors, and is therefore a firm
upper limit on the scatter introduced by any residual systematic error
sources, which we expect to be a few percent at worst.

\subsection{Selection Criterion}

The candidates are selected in the narrow-bands with a detection
threshold of $5 \sigma$, where $\sigma$ is the root-mean-square (rms)
noise determined locally. To reduce the number of foreground [OII] and
[OIII] interlopers, we set the minimum observed equivalent width,
$\hbox{EW} > 80$\AA.  Also, the narrowband flux density should exceed
the broad band flux density at the $4\sigma$ confidence level for a
source to be selected as an emission line source. All these criteria
are designed to keep the numbers of false candidates down to an
acceptable level (Rhoads 2000). An additional condition that the
sources not be detected in B-band image selects against low-redshift
sources, and is roughly equivalent to a photometric redshift. These
criteria yield 157 z=4.5 \lya\ candidates in the three non-overlapping
filters H$\alpha 0$, H$\alpha 8$, H$\alpha 16$.

\subsection{Equivalent width calculation:}

The equivalent width is calculated using narrow and broad band
photometry.  The observed equivalent width is given by $ EW_o =
F_{Ly\alpha}/f_{\lambda} = (W_R \times N - W_N \times R)/(R -N)$,
where $F_{Ly\alpha}$ is the flux of the Lyman$\alpha$ line,
$f_{\lambda}$ is the flux density of the continuum, $W_N$ and $W_R$
are widths of the R band (1568 \AA) and the narrow-bands (80 \AA)
respectively, and $R$ and $N$ are the respective fluxes in the R-band
and the narrow-band in which the source is detected.

The measured equivalent widths are also affected by \lya\ forest
absorption, which can lower both the R-band flux and the line
flux. The correction factor for intergalactic absorption is estimated
using the prescription of Madau 1995.  For the broad band, this is
${\cal A} = 0.58$--$0.68$ for $z=4.57$--$4.37$.  For a \lya\ line that
is symmetric about zero velocity, we find ${\cal A} = 0.64$.  Thus, the
two corrections approximately cancel.  The rest-frame equivalent width
is then given by $EW = EW_O/(1+z)$. The division by (1+z) corrects for
Hubble expansion.

Figure 1 shows the cumulative distribution of rest-frame equivalent
widths after correcting for absorption by the \lya\ forest. The
equivalent widths are derived using equation 1. The continuum is not
detected in approximately half the sources. For such sources we use the
actual measured R flux at the location of the narrowband source, which
makes the equivalent width formally negative in many cases. All such sources
are placed in the last bin of the histogram. The median of the
distribution is then 400 \AA. Figure 1 also shows the cumulative
distribution of EWs if we add 1 and 2 $\sigma$ to the R-band
flux. Some estimate of the uncertainties in the measurement of the EW
can be had from comparing the the three distributions. 

\begin{figure}[ht]
\plotone{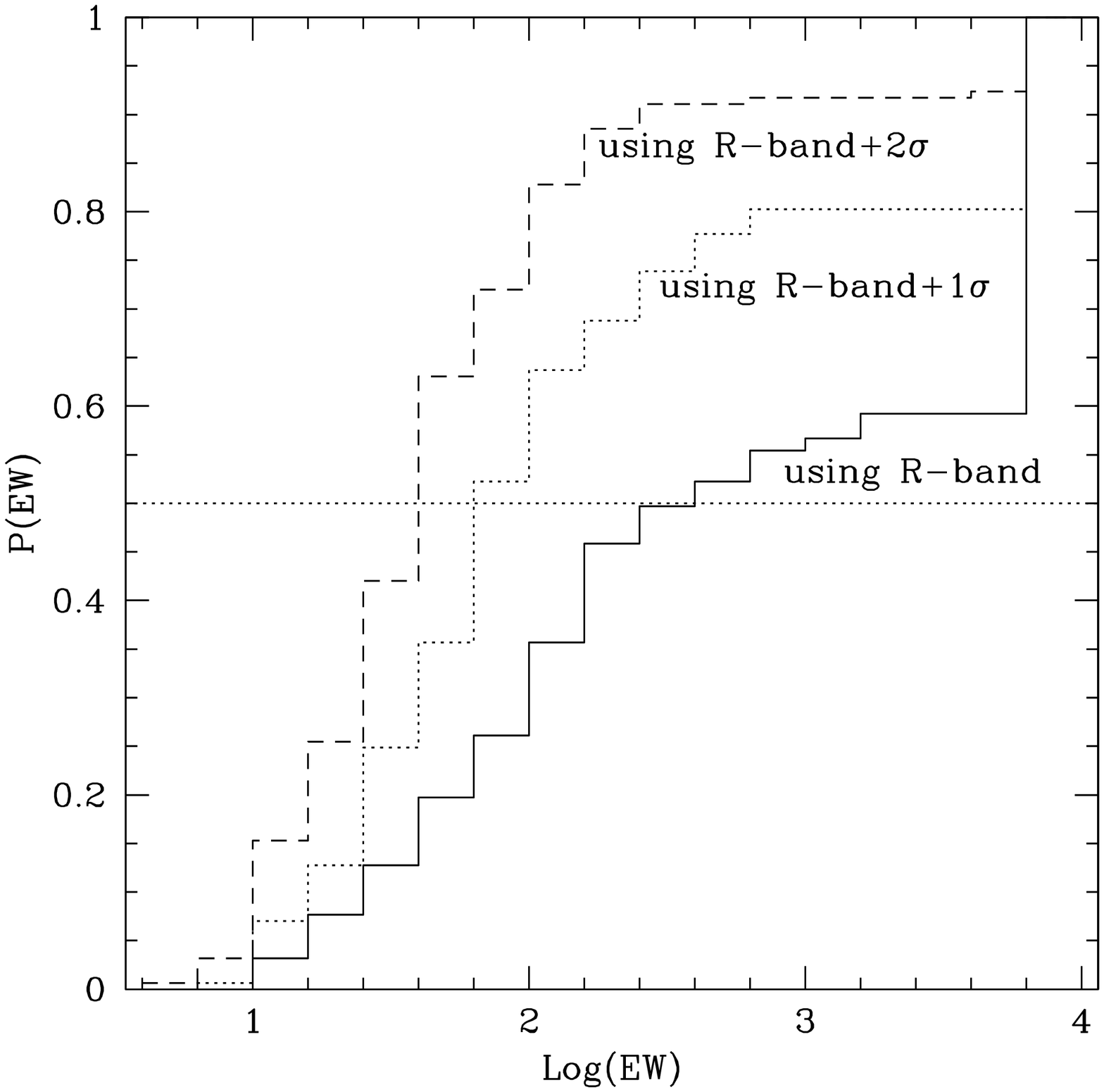}
\caption{The cumulative distribution of \lya\ equivalent widths in the
\lya\ emitter sample. R-band flux is not detected in about half the
sample. The solid line uses measured R-band fluxes to calculate EWs
and sources with negative R-band fluxes are placed in the last
bin. The dotted and dashed lines show the EWs calculated after adding
1 and 2-$\sigma$ flux to the measured R-band flux. More than 60\% of the
sources show equivalent widths $> 240$ \AA, the maximum allowed by
reasonable stellar population models (Charlot \& Fall 1993)}
\label{ewobs}
\end{figure}

\section {Modelling the distribution of equivalent widths}

The equivalent width of the \lya\ line from stellar systems depends on
many factors: the fraction of ionizing photons absorbed by the gas,
the dust content of the interstellar medium (ISM), and the relative
numbers of ionizing and non-ionizing photons in the UV. The last
depends on the slope of the stellar initial mass function (IMF),
metallicity of the stellar population, and age of the system. So we
know the presence of dust decreases the equivalent width of the \lya\
line, while metal poor and more massive, young stars increase
it. Charlot \& Fall (1993) and Kudritzki et al.\ (2000) have calculated
the \lya\ equivalent width for stellar populations with different
properties.

Our approach here is to combine the luminosity function of galaxies at
high redshift with stellar population models and simulate the
observational biases reasonably to model the observed \lya\ emitter
population.  Since we have a reasonably large sample of \lya\
emitters, we can use the distribution of equivalent widths as well as
the total numbers of such objects as constraints on the nature of
these objects.

Three main elements go into modelling the observed number and
distribution of equivalent widths:

(1) The luminosity function of galaxies at z=4.5: Since LAEs are
selected on the basis of a strong line this sample can, and does,
explore the faint end of the luminosity function (Fynbo et al. 2001).
I.e. we can pick up objects with sub-detection level broad-band fluxes
and bright emission lines.  And as long as the luminosity function is
such that there are more faint sources than bright, selection on the
basis of line luminosity will result in bias towards high equivalent
width sources.

To model the effects of such biases and to explore the relation
between line- and continuum-selected high redshift sources, we use the
luminosity function of Lyman Break Galaxies (LBGs) at z=4.0, which is
consistent with the faint end luminosity function from the Hubble Deep
Field (Pozetti et al. 1998, Steidel et al. 1999).  Translating the
luminosity function to a cosmology with $H_0=70, \Omega_M=0.3,
\Omega_{\Lambda}=0.7$ at redshift $z=4.5$, we get a Schecter function
with $\phi_{*}=1\times10^{-3} Mpc^{-3}/\hbox{mag}$ , $\alpha=-1.6$ and
$M_{*}=25.17$.

(2) Stellar population modelling: We estimate the equivalent width of
the Lyman-$\alpha$ line using the stellar population modelling program
``Starburst99'' (Leitherer et al. 1999).


The models were run for continuous star-formation.  The stellar
Initial Mass Function (IMF) is assumed to be a power law with the
exponent ranging between $\alpha=0.5-2.35$, the latter being the
Salpeter law. The metallicity range was from solar to 1/20th
solar. Upper mass cutoff was varied between $M_{upper}=40 - 120
\msun$.  For simplicity we consider three models: Model (A) consists
of continuous star-formation with Salpeter IMF, $Z=1/20$th solar and
$M_{upper}=120 \msun$.  In Model (B) we take the most extreme IMF we
dare and set $\alpha=0.5$.  Model (C) consists of a zero metallicity
stellar population with IMF slope $\alpha=2.35$ whose spectra at the
age of $10^6$ years is derived by Tumilinson \& Shull (2000). The
shape of the age-EW distribution is assumed to be the same as for
model (A) since $\alpha=2.35$ is the same.

The continuum level at 1225 \AA\ is given directly by the models.  The
line strength is derived by assuming that all the ionizing flux is
absorbed by neutral hydrogen and produces 2 \lya\ photons per 3
ionizing photons (case B).  We assume that dust absorption and
resonance scattering have no effect. All this maximizes the \lya\
output from galaxies. The equivalent width of
\lya\ is highest when the galaxies are young and asymptotically approaches
 a steady state value by age $10^7$ years as seen in Figure 2 (see
also Charlot and Fall 1993). For Model C we know the value of EW of
\lya\ at $10^6$ year but not the evolution of the stellar
populations, so we scale the age vs EW curve for model (A) to produce
an \lya\ EW of 1122 \AA\ at $10^6$ years, since the shape of the age
vs EW curve should depend mostly on the IMF slope $\alpha$.

\begin{figure}[ht]
\plotone{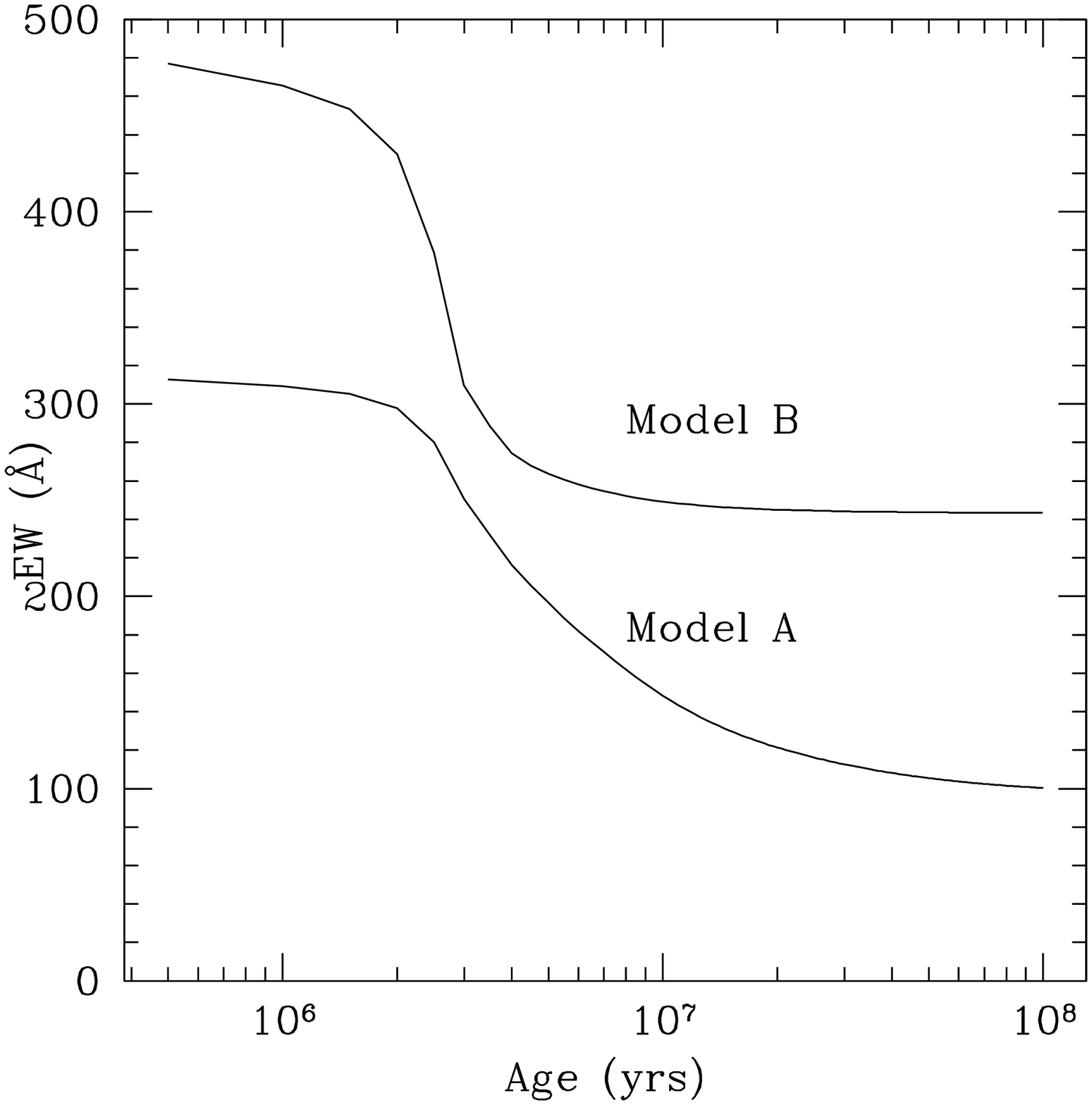}
\caption{The evolution of \lya\ equivalent width as a function of age
for models (A) and (B). Model (C) is assumed to have a similar curve
to (A), except that is is normalized to 1122 \AA\ at age $10^6$ years
(see text for details).}
\label{ewage}
\end{figure}

The \lya\ equivalent width distribution of the galaxy population is
dependent on the age distribution. In this paper we assume as our null
hypothesis that the formation of galaxies is at these epochs is
continuous and uniform in time, which is to say there is no special
synchronization of galaxy formation better than $10^8$ years. The
consequence of this assumption is that there should be 100 times as
many galaxies of age 10$^8$ as 10$^6$ \footnote{ We use age
distribution $0.5\times 10^6-10^8$ years. The upper limit is partly
motivated by the age estimates of Lyman Break Galaxies whose
luminosity function we use and partly by the age of the universe which
is $\sim 10^9$ years at z=4.5}. Therefore the typical equivalent width
of the sample should be the equivalent width of galaxies that are
$10^8$ years old, with a tail extending to the highest equivalent
width for very young galaxies.

(3) Observational selection effects: Since objects are selected on the
basis of having $5\sigma$ flux in the narrow-band and no restriction
in the broad, the error in equivalent widths is dominated by the
measurement errors in broad-band flux. In about half the sample the
continuum is undetected in the broad band R filter at
$2\sigma$. Underestimate of the R-band flux can easily push up the EW
estimates. So we model the effects of R-band measurement errors on the
equivalent width distribution.

Sky noise is the dominant source and is estimated by placing random
apertures on the sky. We take the distribution of fluxes measured in
$10^4$ random apertures that satisfy one of the conditions of \lya\
candidate selection, namely that there be no detectable Bw-band
flux. The distribution of fluxes in such apertures is then taken as
the empirical distribution of errors in the R-band flux and shows
a higher tail than a gaussian distribution.

(4) The Inter Galactic Medium: The R-band flux is reduced due to IGM
absorption blueward of the \lya\ line. We use the prescription by
Madau (1995) to calculate the reduction in R-band flux due to the
absorption of continuum. The \lya\ line flux is also reduced due to
absorption by galactic and inter-galactic gas, leading to the
asymmetric line profile observed in these sources (Rhoads et al. 2000,
Rhoads et al. 2001, Malhotra et al. 2001). Reducing the \lya\ line
flux by a factor of 2 is a reasonable approximation (the IGM alone
absorbs 36\% of the line (Madau 1995)).

Equivalent width (EW$=f_{Ly-\alpha}/f_{\lambda}$) is not the best formalism
to model in a situation where R-band continuum is not detected in half
of the sources and flux density $f_{\lambda}$ is unknown. Instead of
modelling equivalent widths, we work with the ratio of broad-band
to narrow-band fluxes $\Gamma=F_{\lambda}(R)/F_{\lambda}(N)$. With the more uncertain (and small)
quantity in the numerator instead of the denominator, the errors are
much better behaved. Figure 3 shows the histogram of observed $\Gamma$ vs
$\Gamma$ in some of the models. 

Another advantage of this formalism is that in spite of the many
elements that go into reproducing the $\Gamma$  distribution, these elements
are roughly separable in their effects on the distribution. With that
in mind we draw the following robust conclusions\\ 
(1) The total number of \lya\ sources is dependent mostly on the luminosity
function.  We predict the total number of faint galaxies by
extrapolating the LBG luminosity function to lower magnitudes.  Only
7.5-15\% of these galaxies need to show \lya\ emission to match the number of
\lya\ emitters found at z=4.5.\\
(2) The median of $\Gamma$ is unaffected by the noise in R-band flux,
which merely broadens the distribution of $\Gamma$. The median
$\Gamma= 0.0814$, which corresponds to an EW of 430 in the \lya\ line.\\
(3) The width of the $\Gamma$ distribution gives a measure of the
scatter in intrinsic value of $\Gamma$ as well as the flux errors in
the R-band. We find that the latter is the dominant contributer to the
width of the distribution.

\begin{figure}[ht]
\plotone{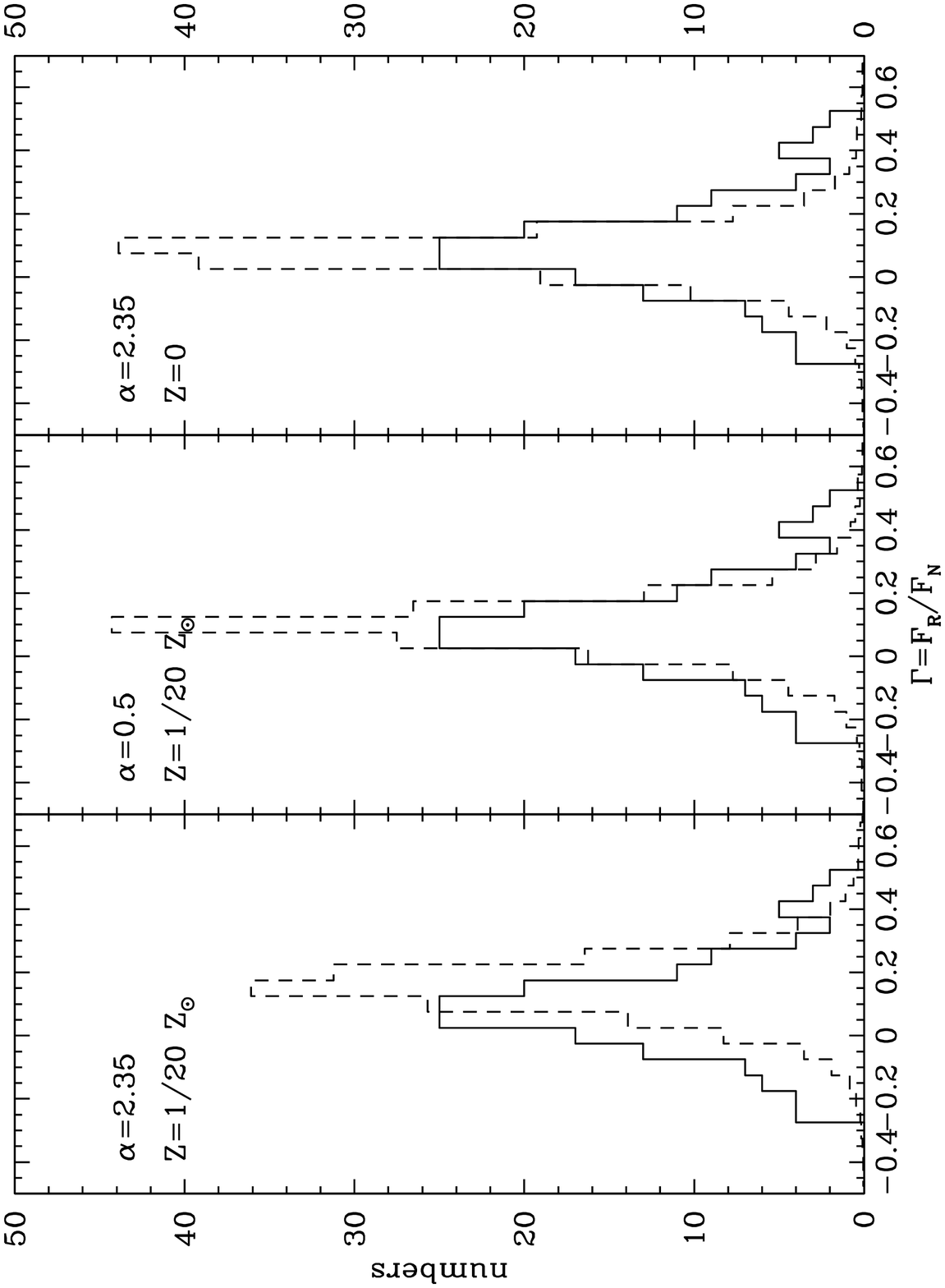}
\caption{The observed distribution (solid histogram) and model
distributions (dashed histograms) of the broad band to narrow band
flux density ratio $\Gamma$ are shown for models (A), (B), and
(C). The models counts are normalised to match the observed numbers of
\lya\ sources. Corrections for IGM absorption have been applied to the
modeled $\Gamma$ values.  Models (B) and (C) fit better than model
(A), though none of the models account fully for the high $\Gamma$
(i.e. low equivalent width) sources.}
\label{models}
\end{figure}

Figure 3 shows how the three models compare with the observed
distribution of $\Gamma$. We see that none of the models fits
perfectly.  Model A clearly produces a higher $\Gamma$ (lower EW) than
observed, and is most clearly disfavored. Models (B) and (C) reproduce
the median $\Gamma$ but do not fully account for the width of the
distribution.  The observed numbers of \lya\ sources requires that a
fraction of faint galaxies show \lya\ emission. This fraction is
15\%, 8.5\% and 7.5\% for Models (A), (B) \& (C)

\section { Discussion and conclusions}

With \lya\ sources we are exploring the faint end of the luminosity
function, because selection on the basis of line luminosity pulls in
sources not easily detected in continuum (Fynbo et al. 2001).  Are we
also exploring the youngest set of galaxies by selecting them on the
basis of their \lya\ emission? In this paper we have tried to answer
that question by studying the equivalent width distribution of these
sources. The median equivalent width of the population is too large to
be explicable by ordinary stellar populations. The distribution of
equivalent widths can be reasonably (though not perfectly) explained
by postulating an IMF that favors more massive stars, or a zero
metallicity population. Neither of these scenarios completely explains
the low EW tails of the distribution.  These tails could be due to
more evolved galaxies where dust effects begin to be observable.
Since the errors in R-band fluxes are responsible for much of the
observed spread in EW, we need deeper R-band images to see the real
spread in EW better.

The number of \lya\ sources is 7.5-15 \% of the numbers expected by
extrapolating the luminosity function of Lyman Break Galaxies. Perhaps
this is a clue that we are only seeing the youngest galaxies and
therefore skewed to-wards high EW of \lya\ line. Why then, do we not
see older galaxies with intermediate EWs in our sample? We are
sensitive to EW$ > 80/(1+z)$. Perhaps dust formation after $10^7$
years quenches the \lya\ line completely, due to resonance
scattering. In another possible scenario, perhaps the early
star-formation in galaxies is episodic, so we only see the galaxies
when the star-formation episode is less that $< 10^7$ years old. That
is also when the galaxies are brighter. This scenario is also
consistent with the blotchy appearances of high redshift galaxies if
the different episodes happen in different parts of a galaxy
(e.g. Colley et al. 1996).

The high equivalent widths can also be a signature of quasar
activity. If so, most of them have to be narrow-line quasars,
otherwise we should see the \lya\ line emission span more than one 80
\AA\ filter. Some of the \lya\  sources could be type II quasars like the ones
found by Norman et al. 2001 \& Stern et al. 2001. Chandra deep field
observations (Rosati et al. 2001) show that there are $\approx 2$--$3 \times
10^3$ x-ray sources per square degree at x-ray fluxes corresponding to
the \lya\ flux limit of our sample.  (The x-ray to \lya\ ratio is fixed
using the Norman et al. object). Thus, if all the \lya\ sources were type
II quasars we would violate the X-ray background constraints by a
factor of 4-5 unless there were no contribution from the X-ray
background from redshifts $z <4$ or $z >5$. This is unphysical and
contradicts the observed distribution of X-ray source redshifts: The
Chandra Deep Field redshift distribution peaks at $z=0.7-0.8$ and
contains no $z > 4$ sources (Gilli et al. in preparation).

To conclude, this paper reports on the EW distribution of \lya\
emitting galaxies at z=4.5 and 5.7. The high median equivalent width
of the distribution requires that the galaxies be very young, have
zero metallicity or stellar IMF of slope $\alpha=0.5$. AGNs can also
produce high EWs. If so, we should be able to detect them with the
current generation of X-ray observatories. High EWs observed in a
large sample obtained by the LALA survey are harder to explain than
individual objects (e.g. Ellis et al. 2001), and therefore pose
questions that will help us understand galaxy and star-formation in 
a young universe.

\acknowledgments 
This work made use of images provided by the NOAO Deep Wide-Field
Survey (NDWFS; Jannuzi and Dey 1999), which is supported by the National
Optical Astronomy Observatory (NOAO).  NOAO is operated by AURA,
Inc., under a cooperative agreement with the National Science
Foundation.
We thank Buell Jannuzi, Arjun Dey, and the rest of the NDWFS team for
making their images public; and Richard Green and Jim De Veny for
their support of the LALA survey. We thank Colin Norman and Tim
Heckman for discussing a number of scenarios to explain these
observations.
JER's research is supported by an Institute Fellowship at The Space
Telescope Science Institute (STScI).
SM's research funding is provided by NASA through Hubble Fellowship
grant \# HF-01111.01-98A from STScI.
STScI is operated by the Association of Universities for Research in
Astronomy, Inc., under NASA contract NAS5-26555.

\end{document}